\pdfoutput=1
\documentclass[a4paper, mathptmx, australian, sort&compress]{aipproc}
\layoutstyle{6x9}
\usepackage{babel}
\usepackage{siunitx}
\usepackage{amsmath}
\usepackage{braket}

\newcommand{\SU}[0]{\mathop{\textrm{SU}}}
\newcommand{\upright}[1]{\mathrm{#1}}

\newcommand{\adjoint}[0]{\dagger}
\newcommand{\conjg}[0]{\ast}
\newcommand{\trsp}[0]{\mathop{\textrm{tr}_\textrm{sp}}}

\newcommand{\ee}[0]{\mathrm{e}}

\begin{document}

\title{Extracting Low-Lying Lambda Resonances Using Correlation Matrix Techniques}
\author{Benjamin J. Menadue}{address={Special Research Centre for the Subatomic Structure of Matter,\\School of Chemistry \& Physics, University of Adelaide, Australia}}
\author{Waseem Kamleh}{address={Special Research Centre for the Subatomic Structure of Matter,\\School of Chemistry \& Physics, University of Adelaide, Australia}}
\author{Derek B. Leinweber}{address={Special Research Centre for the Subatomic Structure of Matter,\\School of Chemistry \& Physics, University of Adelaide, Australia}}
\author{M. S. Mahbub}{address={Special Research Centre for the Subatomic Structure of Matter,\\School of Chemistry \& Physics, University of Adelaide, Australia}}

\begin{abstract}
The lowest-lying negative-parity state of the Lambda is investigated in $(2+1)$-flavour full-QCD on the PACS-CS configurations made available through the ILDG.  We show that a variational analysis using multiple source and sink smearings can extract a state lying lower than that obtained by using a standard fixed smeared source and sink operator alone.
\end{abstract}


\begin{flushright}
ADP-11-08/T730
\end{flushright}
\vspace{-20pt}

\maketitle

\section{Introduction}

The \SI{1405}{\mega\electronvolt} resonance of the Lambda baryon has puzzled researchers for many years. It is the lowest-lying excited state of the Lambda, and yet it has negative parity (a property associated with angular momentum). Moreover, it lies lower than the lowest negative-parity state of the nucleon, even though it has valence strange quarks. It also lacks a nearby spin-orbit partner, with the lowest spin-${3/2}^-$ state being the $\Lambda(1520)$. The internal structure of this resonance has remained a mystery for many years. On one hand, it is regarded as a conventional three-quark state, while on the other it is interpreted as an antikaon-nucleon bound state. 

There have so far been several Lattice QCD studies of this resonance \cite{PhysRevD.67.114506, PhysRevD.68.094505, PhysRevD.74.014504, PTPS.168.598, PoS(LAT2009)108}, however most of these have used the quenched approximation, and very few have managed to identity the mass-suppression associated with the $\Lambda(1405)$. Recent work by the CSSM Lattice Collaboration \cite{PhysRevD.80.054507, PoS(LAT2009)118, SelimNew} has had significant success in isolating the Roper resonance  ($\upright{P}_{11}(1440)$) using correlation matrix techniques together with source and sink smearing, and so we use this method here in an attempt to isolate the otherwise-elusive $\Lambda(1405)$. We use the $(2+1)$-flavour, full-QCD configurations from the PACS-CS collaboration \cite{PhysRevD.79.034503}, made available through the ILDG. In particular, we focus on the configurations with light-quark Hopping parameter $\kappa_{\textrm{u},\textrm{d}} = 0.13770$.

\section{Variational Analysis}

To isolate individual excited states, we use the variational method \cite{Michael198558, Lüscher1990222}, by considering the cross-correlation of operators and diagonalising the operator space. To access $N$ states of the spectrum, we require at least $N$ operators.

The parity-projected, two-point correlation function matrix for $\mathbf{p} = \mathbf{0}$ can be written as
\begin{equation}
G_{ij}^\pm(t) = \sum_{\mathbf{x}} \trsp(\Gamma_\pm \braket{\Omega | \chi_i(x) \smash{\overline\chi}_j(0) | \Omega}) = \sum_{\alpha = 0}^{N-1} \lambda_i^\alpha \smash{\overline\lambda}_j^\alpha \ee^{-m_\alpha t},
\end{equation}
where $\Gamma_\pm$ are the parity-projection operators and $\lambda_i^\alpha$ and $\smash{\overline\lambda}_j^\alpha$ are, respectively, the couplings of interpolators $\chi_i$ and $\smash{\overline\chi}_j$ at the sink and source to eigenstates $\alpha = 0, \ldots, N-1$ of mass $m_\alpha$. The idea now is to construct $N$ independent operators $\phi_i$ that isolate $N$ baryon states $\ket{B_\alpha}$; that is, to find operators $\smash{\overline\phi}^\alpha = \sum_{i=1}^N u_i^\alpha \smash{\overline\chi}_i$ and $\phi^\alpha = \sum_{i=1}^N v_i^{\alpha\conjg} \chi_i$ such that
\begin{align}
\braket{B_\beta, p, s | \smash{\overline\phi}^\alpha | \Omega} &= \delta_{\alpha\beta} \smash{\overline{z}}^\alpha \smash{\overline{u}}(\alpha, p, s),\textrm{ and} \nonumber \\
\braket{\Omega | \phi^\alpha | B_\beta, p, s} &= \delta_{\alpha\beta} z^\alpha u(\alpha, p, s),\label{phidef}
\end{align}
where $z^\alpha$ and $\smash{\overline{z}}^\alpha$ are the coupling strengths of $\phi^\alpha$ and $\smash{\overline\phi}^\alpha$ to the state $\ket{B_\alpha}$. It follows that 
\begin{equation}
G_{ij}^\pm(t) u_j^\alpha = \lambda_i^\alpha \smash{\overline{z}}^\alpha \ee^{-m_\alpha t},\label{Gu=lambdazexp}
\end{equation}
where, for notational convenience, we take the repeated Latin indices to be summed over while repeated Greek indices are not.

The only $t$ dependence in Eq.~\eqref{Gu=lambdazexp} is in the exponential term, so we immediately construct the recurrence relation $G_{ij}^\pm(t) u_j^\alpha = \ee^{m_\alpha \Delta t} G_{ik}^\pm(t+\Delta t) u_k^\alpha$, which can be written as
\begin{equation}
(G^\pm(t+\Delta t))^{-1} G^\pm(t) \mathbf{u}^\alpha = \ee^{m_\alpha \Delta t} \mathbf{u}^\alpha.
\end{equation}
This is an eigensystem equation for the matrix $(G^\pm(t+\Delta t))^{-1} G^\pm(t)$, with eigenvectors $\mathbf{u}^\alpha$ and eigenvalues $\ee^{m_\alpha \Delta t}$.

Similarly, we can construct the associated left-eigensystem equation $\mathbf{v}^{\alpha\adjoint} G^\pm(t) (G^\pm(t+\Delta t))^{-1} = \ee^{m_\alpha \Delta t} \mathbf{v}^{\alpha\adjoint}$, and then Eq.~\eqref{phidef} implies that
\begin{equation}
G^\pm_\alpha(t) := \mathbf{v}^{\alpha\adjoint} G^\pm(t) \mathbf{u}^\alpha = z^\alpha \smash{\overline{z}}^\alpha \ee^{-m_\alpha t}.
\end{equation}
Thus, the only state present in $G^\pm_\alpha(t)$ is $\ket{B_\alpha}$ of mass $m_\alpha$.

We use the ``common'' interpolating fields $\chi_1^\textrm{c}$ and $\chi_2^\textrm{c}$ \cite{PhysRevD.67.114506} that make no assumptions about $\SU(3)$-flavour symmetry, together with three out of the four sink- and source-smearing \cite{Güsken1990361} levels 16, 35, 100, 200. This gives us four bases to construct our $6 \times 6$ correlation matrix; having multiple bases allows us to check for any basis dependence.

\section{Results}

We test each of the four possible operator bases and find that the projected correlation functions agree to within statistical errors. As such, we select the basis $(16, 100, 200)$ as it contains the broadest and most evenly spaced set of smearings. 

\begin{figure}
\includegraphics[width=\linewidth]{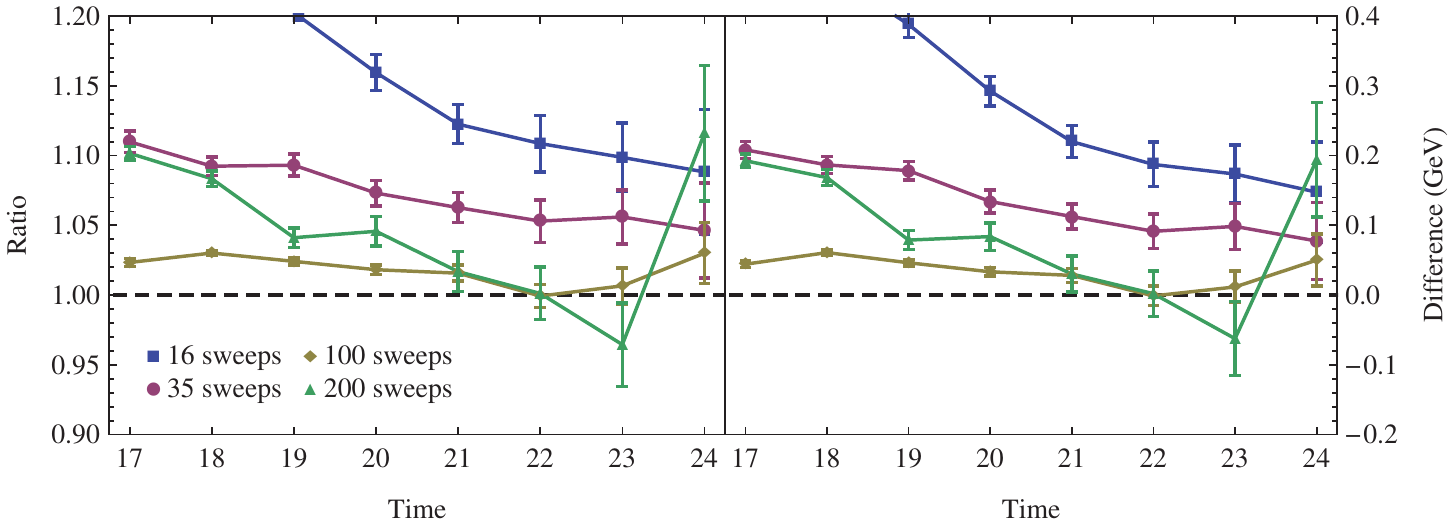}
\caption{\label{comparison}(colour online) Comparisons of the effective mass of the lowest projected negative-parity state $G^-_1(t)$ with that of the diagonal elements $G^-_{ii}(t)$ of the original correlation matrix (those with equal source and sink smearing). The left plot is the ratio $G^-_{ii}(t)/G^-_1(t)$ and the right plot is the difference $G^-_{ii}(t) - G^-_1(t)$. The source is located at $t=16$ and the variational analysis uses $t=18$ and $\Delta t=3$.}
\end{figure}

In order to determine if this approach will help in isolating the low-lying $\Lambda(1405)$, we compare the effective mass of the lowest projected state obtained from $G^-_1(t)$ with that obtained from using a single smeared source and sink without a variational analysis. As we can see from Fig.~\ref{comparison}, the lowest projected state lies lower than any of the diagonal elements. Consequently, we conclude that this method has the potential to enable us to isolate the low-lying $\Lambda(1405)$, and further investigation at the other values of $\kappa$ is warranted in order to calculate a projection of this state to the physical value of $m_\pi^2$.

\bibliographystyle{aipproc}
\bibliography{proceedings}

\end{document}